%% file: paper.tex
\documentclass[twocolumn,showpacs,aps,prd,superscriptaddress,floatfix]{revtex4}
\usepackage{multirow}
\usepackage{dcolumn}
\usepackage{bm}

\newcommand{\BaBarYear}       {07}
\newcommand{\BaBarNumber}     {016}
\newcommand{\SLACPubNumber} {12494}

 \newcommand{\BaBarType}      {PUB}  

\long\def\inst#1{\par\nobreak\kern 4pt\nobreak
    {\it #1}\par\vskip 10pt plus 3pt minus 3pt}

\usepackage[dvips]{graphicx}
\usepackage{epsf}  
\usepackage{epsfig}  
\usepackage{rotating}
\usepackage{color}
\usepackage{colordvi}

\newcommand{\psfile}[3][]{ 
  \begin{center}
    \setlength{\epsfxsize}{#3\linewidth}\leavevmode
    \def\noOpt{}\def\testit{#1}\ifx\testit\noOpt%
      \epsfbox{#2}%
    \else%
      \epsfbox[#1]{#2}%
    \fi
  \end{center}
}

\input babarsym
\input Definitions

\usepackage{relsize}
\usepackage{xspace}
\def\babar{\rm{\slshape B\kern-0.1em{\smaller A}\kern-0.1em
    B\kern-0.1em{\smaller A\kern-0.2em R}}}
\def\pep2{PEP-II}
\def\D0bar{\kern 0.2em\overline{\kern -0.2em D}{\kern 0.1em}\xspace^0}

\def\CP{\ensuremath{C\!P}\xspace}

\begin{document}

{\pagestyle{empty}

\begin{flushleft}
\babar-\BaBarType-\BaBarYear/\BaBarNumber \\
SLAC-PUB-\SLACPubNumber
\end{flushleft}

\title{
        {\mathversion{bold}
         Search for $\Dz-\Dzb$ Mixing Using Doubly Flavor Tagged Semileptonic Decay Modes}
}

\input myPubboard/authors_mar2007

\date{\today}

\input abstract1595

\pacs{13.20.Fc, 12.15.Ff}

\vfill

\maketitle
}

\pagestyle{plain}
%
%
\pagebreak
\input intro1595

\input formalism1595

\input detector
\input analysis1595

%
%
\begin{acknowledgments}
\section{Acknowledgments}
\input myPubboard/acknowledgements

\end{acknowledgments}
\newpage
\bibliography{dmix}

\pagebreak
\input appendix

\end{document}

%% file: Definitions.tex
\providecommand{\bfig}{\begin{figure}[t]}
\providecommand{\efig}{\end{figure}}

\providecommand{\etapr}{\ensuremath{\eta^\prime}}

\providecommand{\KS}{\ensuremath{K_S^0}}

\def\beq{\begin{equation}}
\def\eeq{\end{equation}}
\def\bef{\begin{figure}}
\def\edf{\end{figure}}
\def\ben{\begin{enumerate}}
\def\een{\end{enumerate}}
\def\bear{\begin{array}}
\def\enar{\end{array}}
\def\beqa{\begin{eqnarray}}
\def\eeqa{\end{eqnarray}}

\def\to{\ensuremath{\rightarrow}}

%% file: myPubboard/authors_mar2007.tex
%
\author{B.~Aubert}
\author{M.~Bona}
\author{D.~Boutigny}
\author{Y.~Karyotakis}
\author{J.~P.~Lees}
\author{V.~Poireau}
\author{X.~Prudent}
\author{V.~Tisserand}
\author{A.~Zghiche}
\affiliation{Laboratoire de Physique des Particules, IN2P3/CNRS et Universit\'e de Savoie, F-74941 Annecy-Le-Vieux, France }
\author{J.~Garra~Tico}
\author{E.~Grauges}
\affiliation{Universitat de Barcelona, Facultat de Fisica, Departament ECM, E-08028 Barcelona, Spain }
% ...
\collaboration{The \babar\ Collaboration}
\noaffiliation

%% file: abstract1595.tex
\begin{abstract}

We have searched for $D^0 - \overline{D}^0$ mixing in  
$D^{*+} \to \pi^+ D^0$ decays with $D^0 \to K^{(*)} e \nu$ 
in a sample of $e^+ e^- \to {c\overline c} $ events produced near 
10.58 GeV. 
The charge of the slow pion from charged $ D^* $ decay 
tags the charm flavor at production,
and it is required to be consistent with
the flavor of a fully reconstructed second charm decay in the same event.
We observe 3 mixed candidates compared to 2.85 background events
expected from simulation.  
We ascribe a 50\% systematic uncertainty to this 
expected background rate.
We find a central value for the mixing rate of $ 0.4 \times 10^{-4} $. 
Using a frequentist method, we set
corresponding 68\% and 90\% confidence intervals at 
$(-5.6, 7.4) \times 10^{-4}$ and $(-13, 12) \times 10^{-4}$, respectively.
\end{abstract}

%% file: intro1595.tex
\section{Introduction}
\label{sec:intro}

The $\Dz$ and $\Dzb$ mesons are flavor eigenstates 
which are invariant in
strong interactions,
but are subject to electroweak interactions that permit an 
initial flavor eigenstate to evolve into a time-dependent mixture 
of $\Dz$ and $\Dzb$. 
In the Standard Model (SM), such oscillations proceed through 
both short-distance and long-distance, non-perturbative amplitudes. 
The expected mixing rate mediated by down-type quark box 
diagrams \cite{Datta:1985jx} and di-penguin \cite{Petrov:1997ch} 
diagrams is ${\cal O}(10^{-8}-10^{-10})$, well below the current 
experimental sensitivity of 
${\cal O}(10^{-4}-10^{-3})$ \cite{Burdman:2003rs}. 
The predicted range for non-perturbative, long-distance contributions 
\cite{Burdman:2003rs}
is approximately bounded by the box diagram rate and the current 
experimental sensitivity. 
New physics predictions span the same large 
range \cite{Petrov:2003un}. 
While the presence of a mixing signal alone would not be a clear 
indication of new physics, the current experimental bounds already 
constrain many new physics models.

Because  $\Dz-\Dzb$ mixing has been considered a potential 
signature for new physics, and because $\CP$ violation in such 
mixing would be a signature for new physics, there have been 
many searches for $\Dz-\Dzb$ mixing. 
Typically, these searches use samples of neutral $D$ mesons 
produced as decay products of charged $ D^* $ mesons where 
the charge of the slow pion ($ \pi_s $) produced in association with the 
neutral $D$ meson tags the production flavor of the neutral $D$ meson. 
In semileptonic decays ($\Dz\rightarrow K^{(*)} e \nu$), 
the flavor of the neutral $ D $ meson
when it decays is uniquely identified
by the charge of the lepton.
The signs of the slow pion and lepton charges are the same 
for unmixed decays 
and they differ for mixed decays.
Historically, these two classes of decays are 
denoted as \emph{right-sign} (RS) and \emph{wrong-sign} (WS), 
respectively.

The $B$-factory experiments have searched for $\Dz-\Dzb$ mixing using semileptonic (SL) 
decays, where the initial flavor of the neutral $D$ meson is tagged by the charge of the 
slow pion from a $ D^{* \pm} $ decay. 
The limits on $ r_{\rm mix} $ (defined below) from these 
experiments \cite{Aubert:2004bn, Abe:2005nq}
are listed in Table \ref{PreviousLimitsTable}, 
along with those from recently
published searches for $\Dz - \Dzb$ mixing using hadronic decay 
modes \cite{Asner:2005sz, Zhang:2006dp,
Aubert:2006kt}. 
In the earlier {\babar} SL analysis \cite{Aubert:2004bn}, 
the dominant source of background in the WS signal channel 
originated from RS SL $\Dz$ decays falsely associated with WS slow pion candidates. 
In this analysis we tag the initial flavor of the neutral $D$ meson twice: once using the slow pion 
from the charged $D^*$ decay from which the neutral $D$ decays semileptonically, and once using the 
flavor of a high-momentum $D$ fully reconstructed in the center-of-mass (CM) hemisphere opposite 
the semileptonic candidate. 
Tagging the flavor at production twice, rather than once, highly 
suppresses the background from false WS slow pions but also reduces the signal by more than 
an order of magnitude. 
The \babar\ collaboration has previously used this tagging technique in a measurement of the 
pseudoscalar decay constant $f_{Ds}$ \cite{Aubert:2006sd}. 
We have implemented additional candidate selection criteria to minimize remaining sources of 
background;
the sensitivity of this double-tag analysis is estimated to be about 
the same as that of a corresponding single-tag 
semileptonic analysis for the same dataset.

\begin{table*}[bht]
  \caption{\label{PreviousLimitsTable}
  Limits for $r_{\rm{mix}}$ from earlier measurements in $ e^+ e^- $ experiments.
  }
 \begin{center}
   \begin{tabular}{llrcc} \hline
      Experiment & Decay Mode & Integrated Luminosity & & Upper Limit \\ \hline
      \babar\  \cite{Aubert:2004bn}     &
        $ \Dz \to K^{(*)-} e^+ \nu $ &   $87 \invfb$ &
        \phantom{m} & $ < 42  \times 10^{-4} $ (90\% CL) \\
      Belle  \cite{Abe:2005nq}     &
        $ \Dz \to K^{(*)-} e^+ \nu $ & $254 \invfb$ &
                      &$ < 10 \times 10^{-4} $ (90\% CL) \\
      CLEO \cite{Asner:2005sz}       &
        $ \Dz \to K^0_S \pi^- \pi^+ $ & $9 \invfb$ &
                      &$ < 63  \times 10^{-4} $ (95\% CL) \\
      Belle \cite{Zhang:2006dp}      &
        $ \Dz \to K^+ \pi^- $ &  $400 \invfb$ &
                      &$ < 4.0  \times 10^{-4} $ (95\% CL) \\
      \babar\  \cite{Aubert:2006kt}     &
        $ \Dz \to K^+ \pi^- \pi^0 $ & $230 \invfb$ &
                      &$ < 5.4  \times 10^{-4} $ (95\% CL) \\ \hline
        \end{tabular}
  \end{center}
\end{table*}

%% file: formalism1595.tex

Charm mixing is generally characterized by two dimensionless parameters, $x\equiv\Delta m/\Gamma$ 
and $y\equiv\Delta\Gamma/2\Gamma$, where $\Delta m=m_{2}-m_{1}$ ($\Delta\Gamma=\Gamma_{2}-\Gamma_{1}$) 
is the mass (width) difference between the two neutral $D$ mass eigenstates and $\Gamma$ is the average width.
If either $x$ or $y$ is non-zero, then $\Dz$-$\D0bar$ mixing will occur. 
The decay time distribution of a neutral $D$ meson which changes flavor and decays semileptonically, and thus involves no doubly 
interfering Cabibbo-suppressed (DCS) amplitudes, is \cite{Blaylock:1995ay}:

\begin{equation}
R_{\rm{mix}}(t)\cong R_{\rm{unmix}}(t) \ \frac{x^{2}+y^{2}}{4} \ \left(\frac{t}{\tau_{\Dz}}\right)^{2},
\label{eqn:semileptime}
\end{equation}

\noindent where $t$ is the proper time of the $\Dz$ decay,
$ \tau_{\Dz} $ is the characteristic $ \Dz $ lifetime (= $ 1/ \Gamma )$,
$R_{\rm{unmix}}(t)\propto e^{-t/\tau_{\Dz}}$, and the approximation is valid in the limit of small mixing rates. Sensitivity to $x$ and $y$ individually is lost with semileptonic final states. The time-integrated mixing rate $r_{\rm{mix}}$ relative to the unmixed rate is

\begin{equation}
r_{\rm{mix}} = \frac{x^{2}+y^{2}}{2}.
\label{eqn:mixrate}
\end{equation}

%% file: detector.tex
\section{The {\bf \babar\ } {\bf Detector and Dataset}}
\label{sec:detector}

The data used in this analysis were collected with the \babar\ detector \cite{Aubert:2001tu} 
at the \pep2\ storage ring. The integrated luminosity used here is approximately $344 \invfb$, 
including running both at and just below the $\FourS$ resonance. Charged-particle momenta are measured 
in a tracking system consisting of a five-layer double-sided silicon vertex tracker (SVT) and 
a 40-layer central drift chamber (DCH), both situated in a 1.5-T axial magnetic field. 
An internally reflecting ring-imaging Cherenkov detector (DIRC) with fused silica bar radiators 
provides charged-particle identification. 
A CsI(Tl) electromagnetic calorimeter (EMC) is used 
to detect and identify photons and electrons and measure their energies. 
Muons are identified in the instrumented 
flux return system (IFR).

Electron candidates are identified by the ratio of the energy deposited in the EMC to 
the measured track momentum, the shower shape, the specific ionization measured in the DCH, 
and the Cherenkov angle measured by the DIRC. 
Electron identification efficiency is greater 
than 90\% at all momenta of interest here. 
Pion-as-electron misidentification rates increase from about $0.05 \% $ to 
$ 0.15\% $ from $500 \mevc$ to 3 GeV/$c $. 
Kaon-as-electron  misidentification
rates peak at about $2.5\%$ near $500 \mevc$ and decrease to $0.2\%$ above $800 \mevc$.

Kaon candidates are selected using the specific ionization ($ dE/dx$) measured in the DCH and SVT, 
and the Cherenkov angle measured in the DIRC. 
Kaon identification efficiency is a function 
of laboratory momentum; it is typically 80\% or higher over the range $500 \mevc$ to $3.5 \gevc$, 
with a maximum of about $90\%$ at $2 \gevc$. The pion-as-kaon misidentification rate is 
typically 1\% or less for momenta below $2 \gevc$, rising to about 5\% at $3.5 \gevc$.

%% file: analysis1595.tex
\section{Analysis}
\label{sec:analysis}

The initial selection of semileptonic decay candidates follows the single-tag analysis described
in Ref. \cite{Aubert:2004bn}.
For each $\Dstarp \to \Dz \pi^+; \Dz \to K^{(*)} e \nu$ candidate
(charge conjugation is implied in all signal and tagging modes),
we calculate the  $\Dstarp - \Dz$ mass difference $\Delta M = m(K e \pi ) - m(Ke)$
and the proper lifetime, as well as the output of an event selection neural network (NN).
We then require that a high-momentum $D$ decaying hadronically be fully reconstructed in the opposite hemisphere of the event.
This ensures  that the underlying production mechanism is $e^+ e^- \to c \overline c$ and provides a second production flavor tag.
We implement additional candidate selection criteria based on studies of alternate background samples in data and a Monte Carlo (MC) simulated event
sample (the ``tuning" sample) to reduce various sources of background.
The quark fragmentation in $e^+ e^- \to c \overline c$ MC events
is simulated  using JETSET \cite{Sjostrand:2000wi},
the detector response is simulated via
GEANT4 \cite{Agostinelli:2002hh},
and the resulting events are reconstructed in the same way as are real data.

To minimize bias, we use a MC sample (the ``unbiased'' sample)
disjoint from the tuning sample, to obtain all MC based estimates of efficiencies and backgrounds.
We study this sample,
with effective luminosity roughly equivalent to 603 fb${}^{-1}$
($ \approx 1.75 \times $ data)
 only after all selection criteria and the
full analysis method have been established.
After the expected background rates are determined from the unbiased MC,
we examine the signal region in the data and determine the net
number of observed RS and WS
signal events ($ n_{\rm{RS}} $ and $ n_{\rm{WS}} $).
The measured mixing rate is then determined as $r_{\rm{mix}}=n_{\rm{WS}}/n_{\rm{RS}}$,
corrected for the relative efficiency of the WS and RS signal selection criteria.

\input reco

\subsection{The Hadronic Tagging Samples}
We use the flavor of fully reconstructed charm decays
in the hemisphere opposite the semileptonic signal to
additionally tag the production flavor of the
semileptonic signal, and thus significantly reduce
the rate of wrongly tagged candidates.
We use five hadronic tagging samples.
Three samples explicitly require $\Dstarp$ decays:
$\Dstarp\rightarrow \Dz\pi^+$ where
   $\Dz\rightarrow K^-\pi^+$,
   $\Dz\rightarrow K^- \pi^+ \pi^0$, and
   $\Dz\rightarrow K^- \pi^+ \pi^+ \pi^-$;
while the other two samples are not related to
$\Dstarp$ decays:
  $\Dz\rightarrow K^-\pi^+$ and
  $\Dp\rightarrow K^-\pi^+ \pi^+$.
Candidates from the $\Dstarp $ sample are explicitly excluded from the more
inclusive $\Dz \to K^- \pi^+$ sample to ensure that the tagging samples are disjoint.

\begin{figure*}[thb]
\begin{center}
  \begin{minipage}{1.0\textwidth}
   \epsfig{file=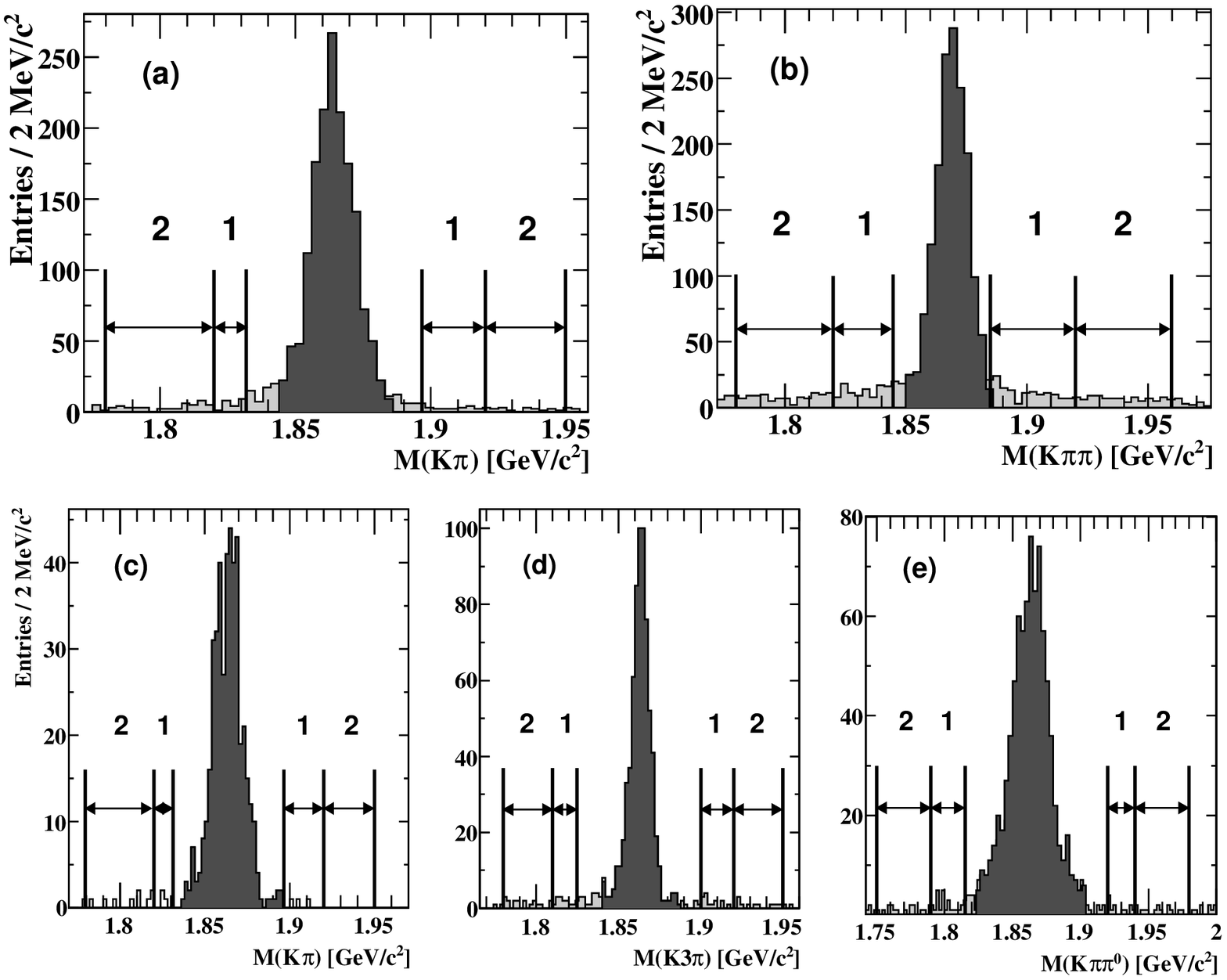,width=0.8\textwidth}
  \end{minipage}
 \caption[]{\label{TaggingFigure}
Invariant mass distributions of hadronic tagging candidates in data events with RS semileptonic
candidates passing the initial semileptonic-side selection.
The dark shaded regions denote hadronic candidates used as tags.
Sideband ``1" events are used to characterize ``false tag" rates and
sideband ``2"
events are used for background normalization in optimizing the hadronic mass selection.
Top row: (a) $\Dz\rightarrow K^-\pi^+$,  (b) $\Dp\rightarrow K^-\pi^+ \pi^+$.
Bottom row: $\Dstarp\rightarrow \Dz\pi^+$ events in final states
(c) $\Dz\rightarrow K^-\pi^+$, (d) $\Dz\rightarrow K^- \pi^+ \pi^+ \pi^-$,
(e) $\Dz\rightarrow K^- \pi^+ \pi^0$.
}
\end{center}
\end{figure*}

The selection criteria for the tagging samples, such as the $ \Delta M $ ranges for the
$ D^* $ modes or the use of production and
decay vertex separation for the $ D^+ $ mode,
vary from channel to channel to balance high purity against high statistics.
Potential criteria are studied using candidate events from a RS sample chosen with
loose requirements on the semileptonic side.
To eliminate candidates from $\BB$ events, we require the CM momentum of the tag-side
$D$ be at least $2.5 \gevc$.
The individual $\Dstarp$, $\Dz$ and $\Dp$ tagging
candidate invariant mass distributions for the final RS sample are shown in Figure~\ref{TaggingFigure}.
The purities of these tagging samples, defined as the ratio of signal
in the selected mass range to the total number of candidates in that range
(the darkly shaded entries in each histogram), vary from about $92\%$
for $\Dz \to K^- \pi^+$ which do not come from $\Dstarp $ to $97.5\%$ for $\Dz \to K^- \pi^+$ from $\Dstarp$.
In the very few events where there are multiple hadronic tag candidates,
we use the tag coming
from the highest purity tagging sample.
We reject events with multiple semileptonic
signal candidates, requiring that one and only one candidate, whether RS or WS,
be present after all tagging and basic semileptonic side selections are imposed.
This requirement rejects approximately $13\%$
of signal candidates and a similar fraction of background
candidates.

\begin{table*}[thb]
\caption{\label{ProgressiveCuts}
Effects of additional semileptonic side selection criteria. Approximate cumulative acceptance rates in the double-tag unbiased MC for signal and WS
background as additional selection criteria are applied, relative to acceptances following the initial semileptonic-side selection. The signal
acceptances are the same for RS and WS signal samples except for the decay time cut where the entry is that for the WS sample.}
\begin{center}
\begin{tabular}{ccc} \hline
Criterion & Signal Retained & WS Background Retained \\ \hline
$ e^{\pm} $ conversion and Dalitz pair veto & 100\% & 82\% \\
$ \pi_s $ $\dedx $ cut                      &  85\% & 66\% \\
$ \pi_s $ $ p_T $ and $ p_L $ selection     &  72\% & 36\% \\
$ m(Ke) > 0.8 \gevcc$                       &  71\% & 30\% \\
$ ( M(Ke), \Delta M ) $ kinematic cut       &  70\% & 20\% \\
600 $ < t < $ 3900 fs                       &  55\% & 10\% \\ \hline
\end{tabular}
\end{center}
\end{table*}

\subsection{Additional Semileptonic Side Selection Criteria}
Double-tagging the production flavor of the neutral $D$ mesons effectively eliminates the
WS background due to real semileptonic $D$ decays paired with false slow pions from putative $\Dstarp $ decay.
From studies of background events in the tuning MC sample,
we find that kaon and electron
candidates are almost always real kaons and electrons, respectively, with correctly assigned charges.
We also find that many fake slow pion candidates are electrons produced as part of conversion pairs,
or Dalitz decays of $\piz$ or, to a much lesser extent, $\eta$ mesons.
These processes also contribute background tracks to the pool of electron candidates
used to create $Ke$ vertices combinatorically.
We consequently implement selection criteria to reject tracks
which may have
originated in such processes by requiring that neither an electron nor slow pion candidate
form a conversion pair when combined with an oppositely charged track treated as an electron
(whether identified as such or not). We further require that electron candidates not form
a $\pi^0$ candidate when combined with a photon candidate and an oppositely charged track treated as an electron.
(After applying all event selections, we find no contribution in the tuning MC sample from $\eta$ Dalitz decays.)
Rejecting photon conversions and Dalitz decays reduces the total RS and WS backgrounds by
about $20\%$ each, and has a negligible effect on signal efficiency.

To reduce backgrounds from kaons misidentified as electrons,
we require that the laboratory momentum of electron candidates be greater than $600 \mevc$.
This reduces the signal efficiency by about $15\%$ and the background rate by about $35\%$.
To further reduce the number of electrons that are considered as slow pions, we veto tracks
where $\dedx$ in the SVT is consistent with that of an electron.
This reduces the signal efficiency by about $15\%$ and the background rate by about $25\%$.

We study kinematic distributions that discriminate between signal and
background using data and MC events with two fully reconstructed hadronic
decays of charm mesons.
As a result, we require that the slow pion CM longitudinal momentum
(along the axis defined by the direction opposite the
tagging $D$'s CM momentum) lie in the range $150-400 \mevc$ and its
transverse momentum
be less than $80 \mevc$.
This reduces the background by approximately $40\% $ and the signal efficiency by
approximately $15\%$.

\begin{figure}[tb]
\begin{center}
\begin{minipage}{.5\textwidth}
\epsfig{file=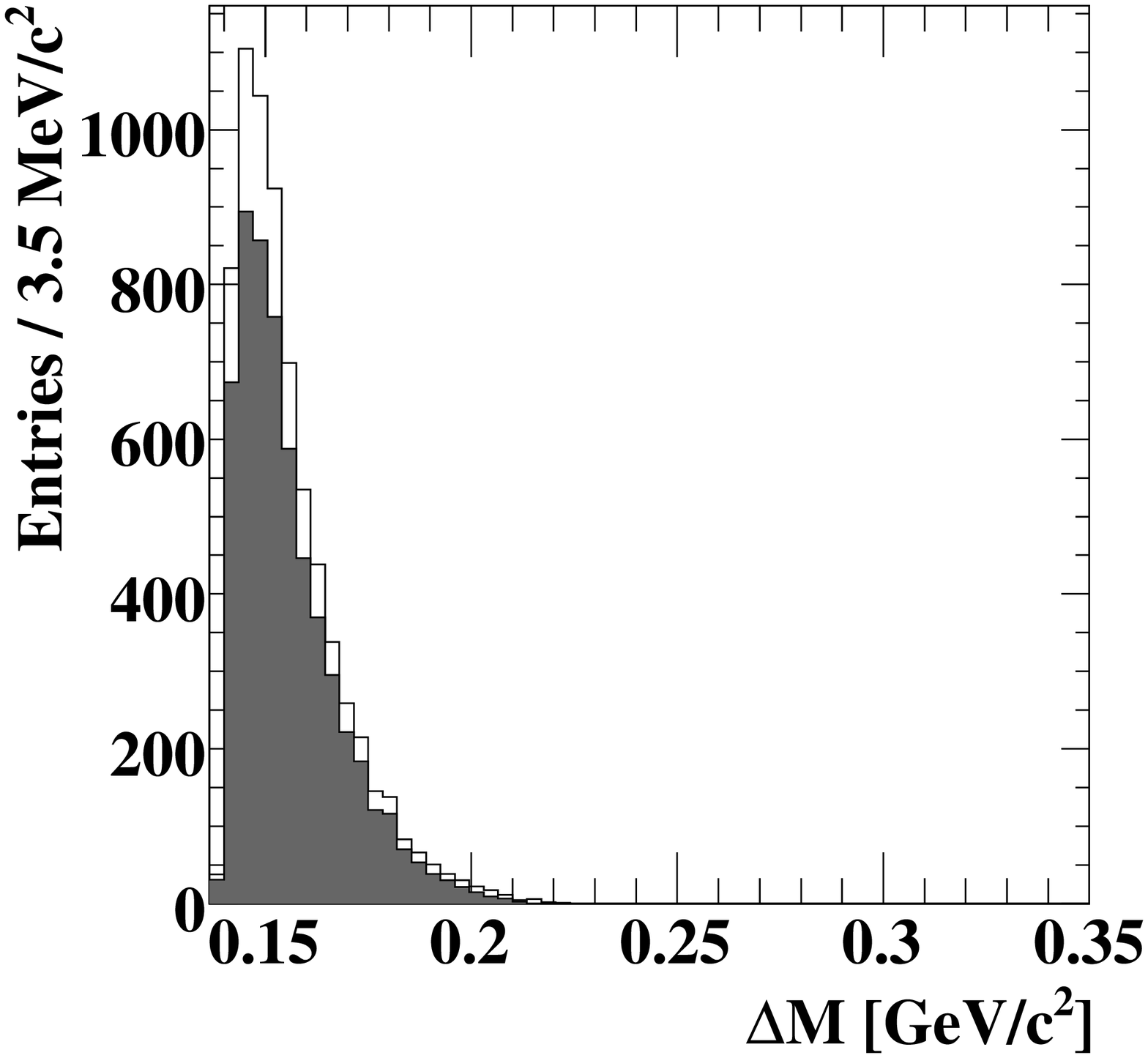,width=0.9\textwidth}
\end{minipage}
\caption[]{\label{fig:dmRSsgnl}
RS signal $\Delta M$ distribution in unbiased MC scaled to the luminosity of the data before (line) and after (solid) applying the double-tag kinematic
selection.}
\end{center}
\end{figure}

We require that the electron-kaon invariant mass be greater than $800 \mevcc$;
this reduces the signal efficiency by a few percent while it reduces the
background rate by approximately $20\% $.
When we count the final number of signal candidates, we also require that
$ (M(Ke), \Delta M )$ lies inside the kinematic boundary
expected for
$ D^{*+} \to D^0 \pi^+; \ D^0 \to K e \nu $ decays where the neutrino
momentum is ignored.
This has essentially no effect on signal efficiency, and reduces WS backgrounds by about $35\% $.
The cumulative effects of the additional semileptonic side selection criteria are summarized
in Table \ref{ProgressiveCuts}.
(The selection for electron momentum $>600 \mevc$ is applied
prior to calculating the acceptances
listed in the table.)
The effects of these additional selection criteria on
$ \Dz \to K^{*} e \nu $  events
are reasonably consistent with those for
$ \Dz \to K e \nu $ events.
The combination of the slow pion longitudinal and transverse momenta and the $(M(Ke), \Delta M)$
selections will hereinafter be referred to as the ``double-tag kinematic selection''.
Figure~\ref{fig:dmRSsgnl} shows the $ \Delta M $ distributions of signal events in unbiased MC scaled
to the luminosity of the data both before and after imposing the double-tag kinematic selection.
The marginal efficiency resulting from applying these last selection criteria to
signal events is $84 \pm 1\% $.

\begin{figure}[tb]
\begin{center}
\begin{minipage}{.5\textwidth}
\epsfig{file=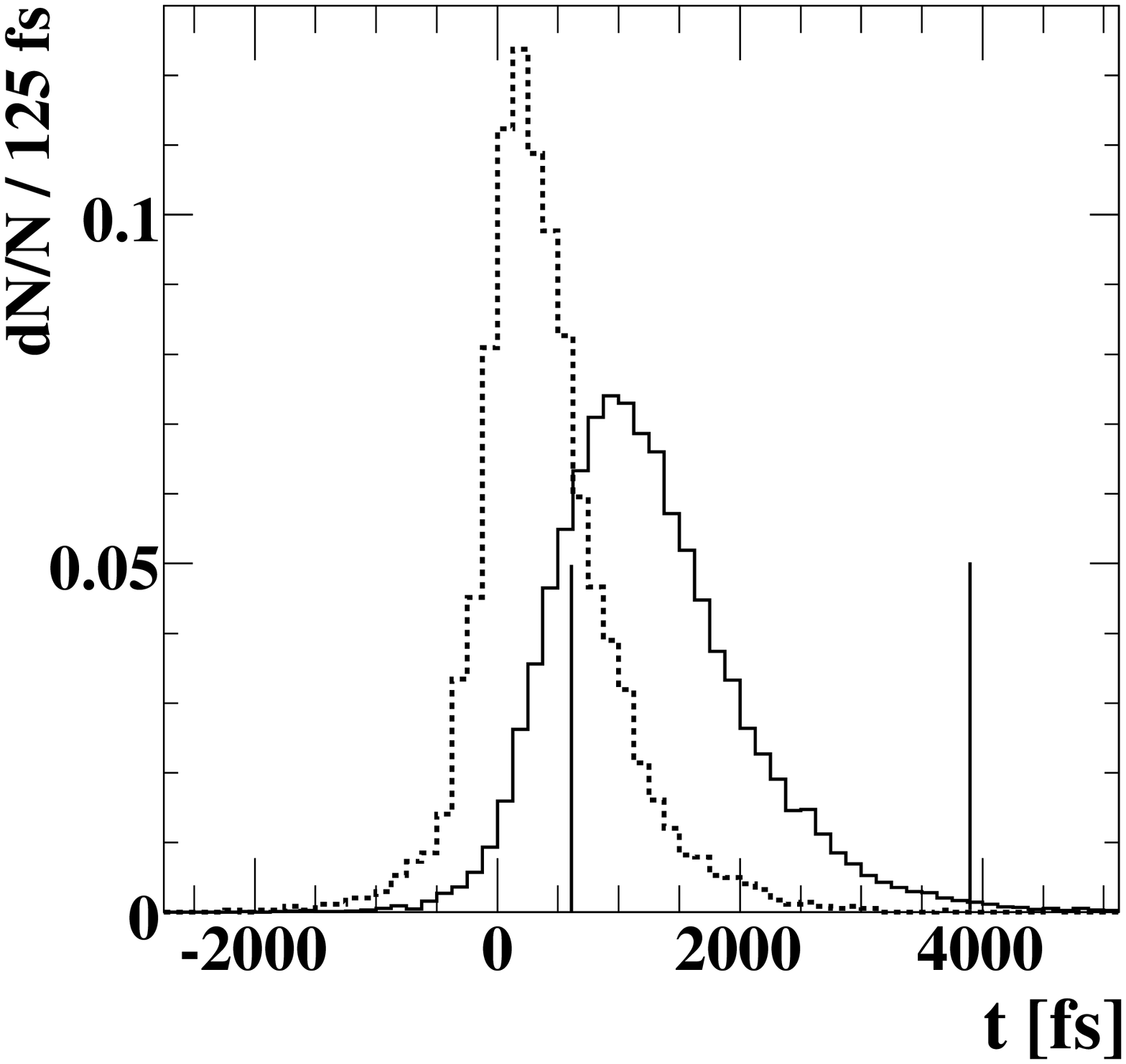,width=0.9\textwidth}
\end{minipage}
\caption[]{\label{fig:taursws}
Normalized RS (dashed) and WS (solid) reconstructed simulated signal
lifetime distributions.
The solid vertical lines mark the range for the selection of the WS events.}
\end{center}
\end{figure}

The decay time distributions of the RS and WS signals should differ,
as shown in Equation \ref{eqn:semileptime}.
The RS sample is produced with an exponential decay rate, while the WS sample should
be produced with the same exponential rate
modulated by $ t^2 $.
Figure~\ref{fig:taursws} shows the normalized lifetime distributions
for reconstructed simulated RS and WS signal events passing the final tag and signal-side selection.
To improve sensitivity, we select only WS candidates with measured lifetimes between 600 fs $ ( \approx 1.5 \tau_{\Dz} ) $ and 3900 fs
$ ( \approx 9.5 \tau_{\Dz} ) $, which accepts approximately $80\%$
of signal and less than 30\% of background.
Because the RS signal-to-background ratio is comparatively very large,
we accept RS candidates
across the full range shown in Figure~\ref{fig:taursws}.
This WS/RS relative efficiency
has a 2\% systematic uncertainty due to imperfect knowledge of the decay
time resolution function.
This  is determined from changes in
the WS/RS efficiency observed when varying the signal resolution function
according to the difference between resolution functions observed in RS data and MC samples.

\begin{figure}[tb]
\begin{center}
\begin{minipage}{.5\textwidth}
\epsfig{file=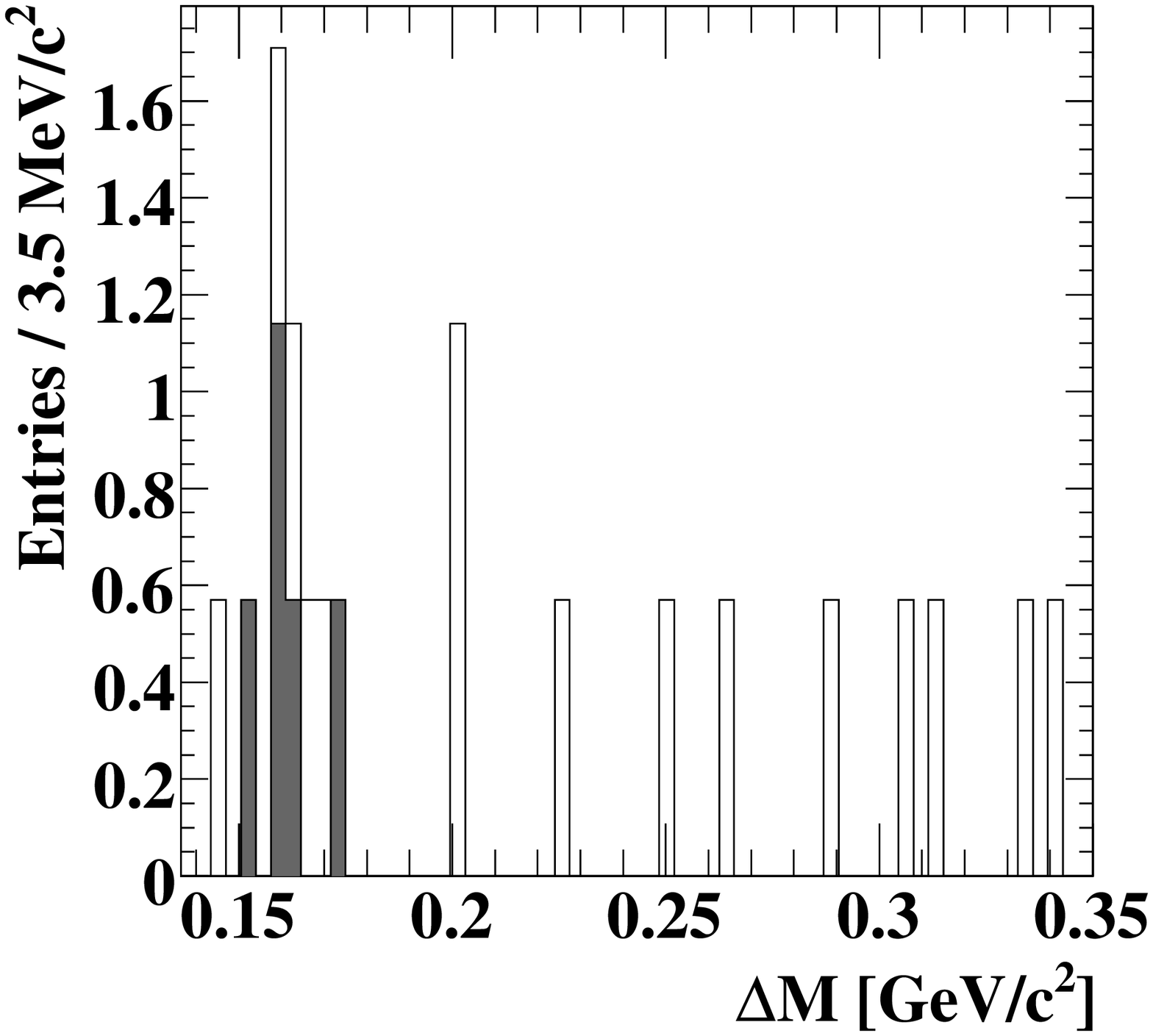,width=0.9\textwidth}
\end{minipage}
\caption[]{\label{fig:dmWSbkgd}
WS $\Delta M$ distribution for background events passing the WS decay time selection in unbiased MC scaled to the luminosity of the data before (line)
and after (solid) applying the double-tag kinematic selection.}
\end{center}
\end{figure}

Figure~\ref{fig:nn}(b) shows the NN event selector output for RS signal,
RS backgrounds and WS backgrounds in the unbiased MC sample
passing the additional semileptonic side selection criteria (scaled to the
luminosity of the data).
The effectiveness of the additional semileptonic-side criteria in suppressing WS
backgrounds while simultaneously retaining good signal efficiency can be seen by
comparing Figures~\ref{fig:nn}(a) and (b).
Figure~\ref{fig:dmWSbkgd} shows the $\Delta M$ distribution of WS backgrounds passing
the decay time selection in unbiased MC scaled to the luminosity of the data both
before and after the double-tag kinematic selection.
A total of 2.85 background candidates, the sum of the luminosity-scaled events in the
solid histogram shown in the figure, is expected after all event selection criteria are applied.

\subsection{Measuring Signal Yields}

To determine the mixing rate,
we first determine
the number of RS signal candidates by fitting the RS
$\Delta M$ distribution,
as described in detail below.
We then estimate the expected rate of WS background events
in the signal region of the
data from the unbiased MC sample.
Using several background control samples drawn from both data and MC,
we estimate how well MC events describe real data events.
Using a statistical procedure with good frequentist coverage,
we combine
the number of candidates observed in the WS sample,
the expected background rate,
and the estimated systematic uncertainty in the expected background rate
to obtain a central value for the mixing rate and
68\% and 90\% confidence intervals.
This procedure is described in detail in the appendix.

We extract the number of RS signal events from the $\Delta M$ distribution of the
RS sample selected without the double-tag kinematic selection using an
extended maximum likelihood fit.
The likelihood function includes probability density functions
(PDF's) for the signal,
the background events which peak in the signal region,
and the combinatorial background.
The PDF for each event class is assigned using the functional
forms described in Ref. \cite{Aubert:2004bn}.
The shape parameters for the combinatoric background are determined using the following
technique: $\Dz$ signal candidates in the data from one event are combined with $\pi_s$
candidates from another event to model the shape of this PDF. Based on MC studies,
the shape of the peaking $\Delta M$ background is assumed to be the same shape as the signal.
Its relative level is also determined from MC studies.
The shape parameters of the signal PDF, as well as the number of RS signal events and
the number of combinatorial background events, are then obtained from the likelihood fit of the data.

\begin{figure}[tb]
\begin{center}
  \begin{minipage}{.5\textwidth}
   \epsfig{file=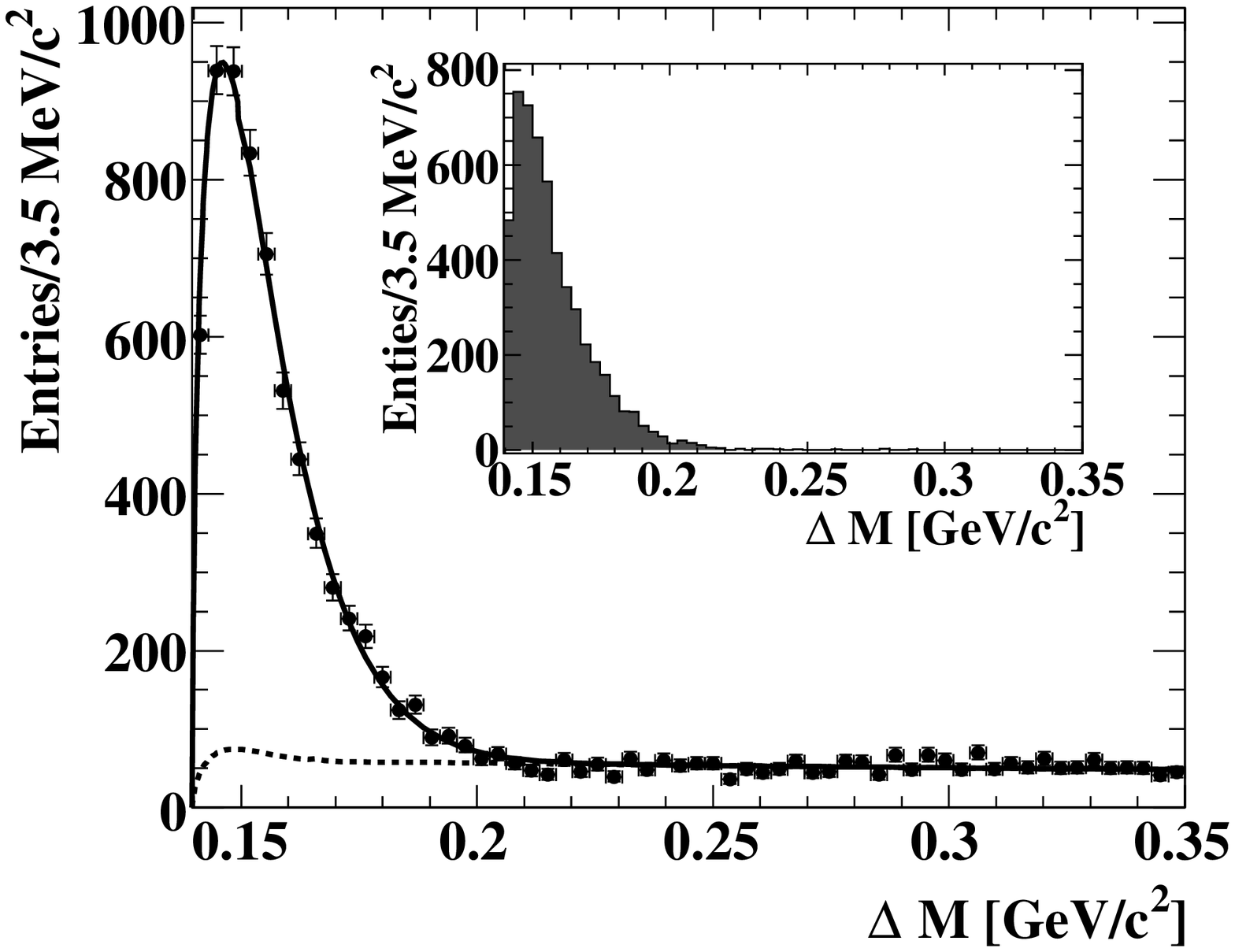,width=0.9\textwidth}
  \end{minipage}
 \caption[]{\label{FinalRSFig}
RS data $\Delta M$ distribution.
The main plot shows the RS data (points) before imposing the double-tag kinematic selection,
and the projections of the total fit PDF (solid line) and the background PDF (dashed line).
The inset plot shows the RS $\Delta M$ distribution after the double-tag kinematic selection
criteria are applied.}
\end{center}
\end{figure}

The main plot in Figure~\ref{FinalRSFig} shows the $\Delta M$ fit of the RS data
before applying the double-tag kinematic selection, with the signal and background
contributions overlaid.
The fitted RS signal yield in this sample is $5748 \pm 90$ events, with $\chi^2=77$ for 60 bins,
where six parameters are determined from the fit.
The inset plot of Figure~\ref{FinalRSFig} shows the  RS data
$\Delta M$ distribution
after the double-tag kinematic selection is imposed. As noted above, the efficiency
of this selection is $0.84 \pm 0.01$, giving a final RS signal yield of $4780 \pm 94$,
which is used as the normalization in calculating the mixing rate.

\begin{figure}[tb]
\begin{center}
  \begin{minipage}{.5\textwidth}
   \epsfig{file=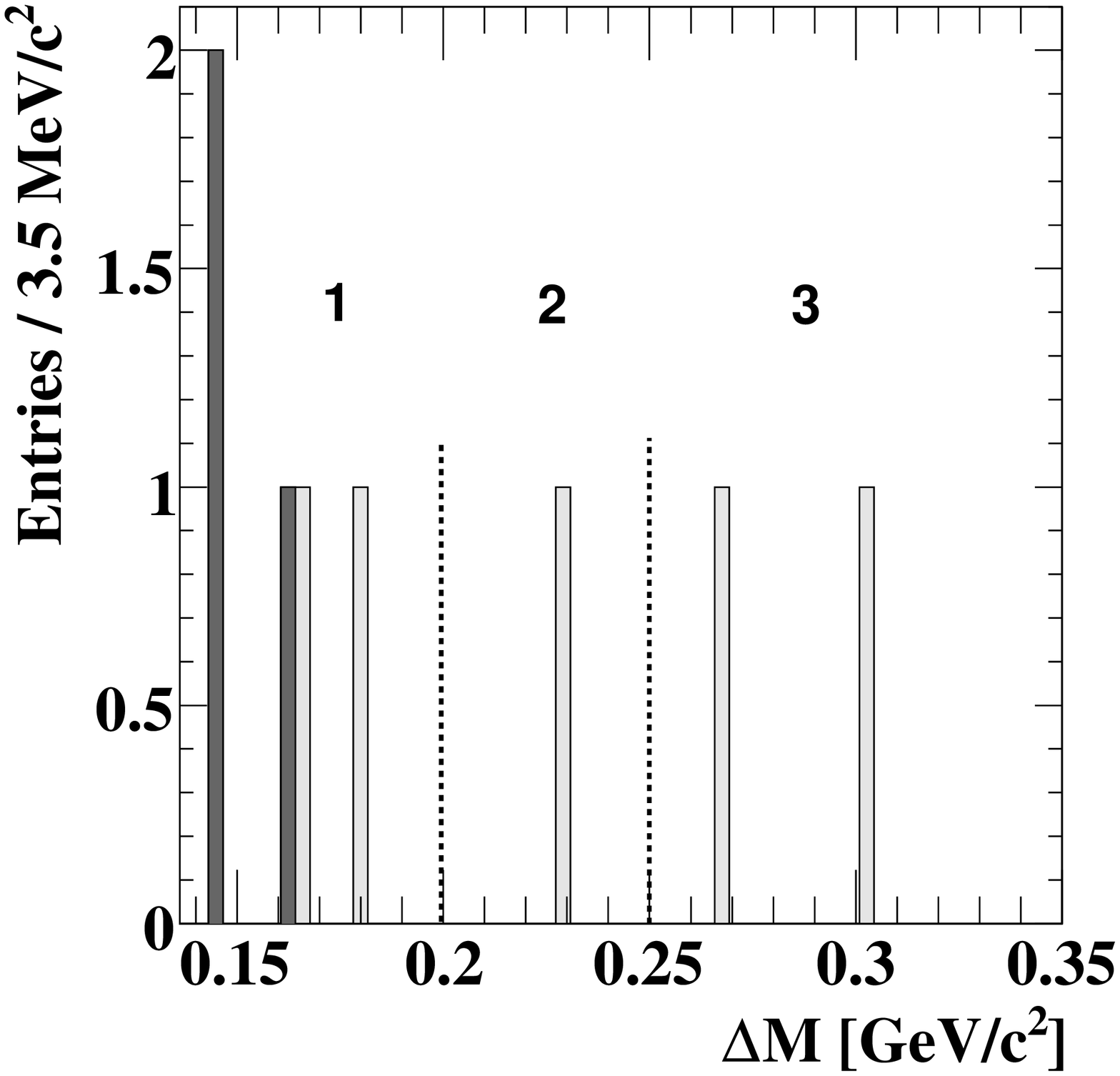,width=0.9\textwidth}
  \end{minipage}
 \caption[]{\label{FinalWSFig}
WS data $\Delta M$ distribution.
The dark histogram shows WS events in the data passing all event selection.
The light histogram shows WS events passing all selections except the double-tag
kinematic selection.
Region ``1'' is the signal region, ``2'' is the near sideband, and ``3'' is the far sideband.
}
\end{center}
\end{figure}

To determine the number of WS mixed events, we consider three regions of $\Delta M$:
the signal region, $\Delta M \le 0.20 \gevcc$;
the near background region, $0.20 < \Delta M \le 0.25 \gevcc$;
and the far background region, $0.25 < \Delta M \le 0.35\gevcc $.
These $  \Delta M $ ranges are shown in Figure ~\ref{FinalWSFig},
and are respectively labeled ``1", ``2" and ``3" in the plot.
To avoid potential bias,
we examine neither the signal region nor the near background region
in the WS data sample until all of the selection criteria and the procedure for calculating
confidence intervals are determined.
The WS signal region may contain both signal and background events after applying the final
event selection critera.
As discussed above, we determine the expected number of background events from the unbiased
MC sample: we observe 5 events,
which scales to  2.85 for the luminosity of the data.
To estimate the possible non-$c \overline c$
background rate, we also examine events which satisfy the semileptonic-side selection criteria
but fail the tagging-side criteria because the mass of the hadronic $D$ candidate falls outside
the accepted window.
Since we had examined the data events in the ``far" sidebands
(sidebands ``2") of Figure~\ref{TaggingFigure} while optimizing hadronic
side selection criteria,
we also examine those in the ``near sidebands (sidebands ``1")
to estimate the number of these ``false tag" events:
we find no WS candidates in the near or far $\Delta M$ sideband regions in either the data or
unbiased MC sample.
Given the agreement between data and the unbiased MC sample, we determine the central value of
the number of WS signal events by subtracting the luminosity-scaled number of unbiased MC WS
background events in the signal region from the number of candidates observed in the data there.

The dark shaded entries in Figure~\ref{FinalWSFig} denote the $\Delta M$ distribution of WS
candidates in the data after all event selection, where we observe $3$ WS candidates
in the signal region and none in the sideband regions.
Given the expected WS background of 2.85~events shown in the solid histogram of
Figure~\ref{fig:dmWSbkgd}, we calculate a net WS signal yield of 0.15 events.
We discuss below the total error associated with the estimated number
of WS background events.

\subsection{Systematics and Confidence Intervals}

To calculate confidence intervals for the number of mixed events observed,
we first determine a systematic uncertainty associated with the
WS background estimate.
To do this,
we compare 10 background control samples in data with the
corresponding MC samples.
The results of this comparison are shown in
Table \ref{ProxyBackgroundTable}.
The first line compares the number of WS events observed in the
far background region of the data and the tuning MC sample.
The second line compares the same numbers for the data and for
the unbiased MC sample.
The remaining table entries compare the number of events observed
in two types of doubly-charged (DC) background samples obtained
from data with those observed from the same sources in
unbiased MC events.
In both of the DC background samples, the kaon and the electron have
the same charge sign, and are reconstructed exactly as neutral
$ K e $ vertex candidates are, except for the differing charge correlation.
In those additionally labeled WS, the slow pion has the same charge
as the kaon, while in those additionally labeled RS,
the slow pion has the opposite charge.

\begin{table*}[tb]
\caption{\label{ProxyBackgroundTable}
Comparison of MC and data background yields. The doubly-charged (DC) MC entries refer to MC event samples disjoint from those used to optimize event
selection. The ``kinematic selection" refers to the double-tag kinematic selection.
}
  \begin{center}
        \begin{tabular}{cccccc} \hline
          Entry & Data Sample & $\Delta M$ Range (\gevcc) &
          kinematic selection  & Data & \phantom{mm} MC \phantom{mm} \\ \hline
          1 & WS, tuning MC & $0.25 \le \Delta M \le 0.35$   & no & $2\pm 1.4$ & $2.1 \pm1.5$ \\
          2 & WS, unbiased MC & $0.25 \le \Delta M \le 0.35$   & no & $2\pm1.4 $ & $3.4 \pm1.4$ \\
          3 & DC, RS & $\Delta M \le 0.20$   & yes & $37 \pm 6$ & $40 \pm 5.1$ \\
          4 & DC, WS & $\Delta M \le 0.20$   & yes & $36 \pm 6$ & $51 \pm 5.8$ \\
          5 & DC, RS & $\Delta M \le 0.20$   & no & $42 \pm 7$ & $47 \pm 5.5 $\\
          6 & DC, WS & $\Delta M \le 0.20$   & no & $ 55 \pm 8$ & $64 \pm 6.5$ \\
          7 & DC, RS & $0.20 < \Delta M \le 0.25$ & no   &
          $20 \pm 5$ & $ 24 \pm 3.9$ \\
          8 & DC, WS & $0.20 < \Delta M \le 0.25$ & no   &
          $13 \pm 4$ & $ 19 \pm $ 3.5\\
          9 & DC, RS & $0.25 \le \Delta M \le 0.35$ & no   & $20 \pm 5$ & $31 \pm 4.5$\\
          10 & DC, WS & $0.25 \le \Delta M \le 0.35$ & no   & $23 \pm 5$ & $18 \pm 3.4$ \\ \hline
        \end{tabular}
\end{center}
\end{table*}

Ignoring the correlations between entries 3,5 and 4,6 in
Table \ref{ProxyBackgroundTable}, we estimate the consistency
between the data and MC samples by calculating a
summed $\chi^2 $ for all the entries:

\begin{eqnarray}
\chi^2 (\mbox{data, MC}) = \sum\limits_{i=1}^{10} \biggl [\frac{(x_i^{data}-x_i^{MC})^2} {(\sigma_i^{data})^2 + (\sigma_i^{MC})^2} \biggr] = 11.4
\end{eqnarray}

The value $\chi^{2}=11.4$ is consistent with 1 per degree of freedom.
Taken together, these observations indicate that the MC estimate for
the background rate in the signal region of the WS sample is reasonably accurate.
We conservatively assign the largest discrepancy between the data and MC rates,
50\%, as the systematic uncertainty associated with the ratio between the MC
estimate of the background rate and its true value.

To determine confidence intervals for the number of WS mixed events,
we adapt a suggestion made in Ref. \cite{Porter:2003ui}.
The complete statistical procedure is described in detail in the appendix;
it is summarized here.
We start with a likelihood function, $ {\cal L} (n, n_b; s, b) $, for the number of events
observed in the signal region of the WS data sample, $ n $, and the corresponding number
observed in the MC sample, $ n_b $. $ {\cal L} (n, n_b; s, b) $ depends upon the true signal
rate $ s $ and the true background rate $ b $ in the signal region, and also accounts for the
systematic uncertainty in the ratio of the true background rate in data to that estimated
from MC.
The value of $ (s, b ) $ which maximizes the likelihood function,
$ {\calL}_{\rm max} $, is denoted by $ (\hat s , \hat b ) $.
As one expects naively, $\hat b $ is equal to $ n_b $ times
the ratio of data and MC
luminosities while $\hat s  = n - \hat b $. We then search for the values of $ s $
where  $ - {\rm ln} {\cal L} (s) $ changes by 0.50 [1.35]; here  $ {\cal L} (s) $
denotes the likelihood at $ s $ maximized with respect to $ b $.
The lower and upper values of $ s $ which satisfy this condition define the
nominal 68\%  [90\%] confidence interval for $ s $.
As discussed in the appendix, for the range of parameters relevant for this analysis,
the confidence intervals produced by this procedure provide frequentist coverage which
is accurate within a few percent.

\subsection{Final Results and Conclusion}

We observe 3 candidates for $\Dz-\Dzb$ mixing,
compared to  2.85 expected background events, where
we ascribe a 50\% systematic uncertainty to this
expected background rate.
We find the central value for the number of
WS signal events to be 0.15, with 68\% and 90\% confidence intervals
$ (-2.2, 2.8) $
and $ (-5.2, 4.7) $, respectively.
Accounting for the ratio of WS and RS signal efficiencies due to the cut on the measured
WS decay time ($0.80 \pm 0.02$), we find the central value of
$ r_{\rm mix} $ to be $ 0.4  \times 10^{-4} $,
with 68\% and 90\% confidence intervals
$ (-5.6, 7.4) \times 10^{-4}$ and $ (-13, 12) \times 10^{-4} $, respectively.
We ignore variations in the RS yield due to statistical error and systematic effects in the RS
fit as they are negligible relative to the statistical errors associated with the WS data and MC rates as well as the 50\% systematic error assigned to
the ratio of MC and data WS background rates.

The sensitivity of this double-tag analysis is comparable to that
expected for a single-tag analysis, see Table \ref{PreviousLimitsTable}.
Future analyses should be able to combine these two approaches to significantly improve overall sensitivity to charm mixing using semileptonic final
states. Improved methods for reconstructing and selecting semileptonic signal candidates, the use of more hadronic tagging modes, and the additional
use of semi-muonic decay modes
may allow semileptonic charm mixing analyses to approach the $ r_{\rm mix} $ sensitivity of analyses using hadronic final states.

%% file: reco.tex
\subsection{Reconstruction and Selection of Semileptonic Signal Candidates}

Semileptonic signal candidates are selected by reconstructing the decay chain
$\Dstarp \rightarrow \pi^{+} \Dz, \Dz \rightarrow K^{(*)}e\nu$.
There are no essential differences for this analysis between the
$K$ and $K^{*} \to K^{\mp} \pi^0 $
modes, either theoretically or empirically, and thus no attempt is made to
reconstruct the $K^{*}$ --- its charged $K$ daughter is treated as if it were a
direct daughter of the $\Dz$. Approximately 11\% of signal candidates
accepted in the initial selection of semileptonic canidates \cite{Aubert:2004bn}
are in
the $K^{*}$ mode.

Identified $K$ and $e$ candidates of opposite charges are combined to create
neutral candidate $D$ decay vertices. Only candidates with vertex fit
probability $>$ 0.01 and invariant mass $<$ 1.82~GeV/$c^{2}$ are retained.
This requirement is imposed to exclude all hadronic two-body $\Dz$ decays.
The average PEP-II interaction point (IP), measured on a run-to-run basis
using Bhabha and $\mu^{+} \mu^{-}$ events, is taken as the production
point of $\Dz$ candidates.
The  $\Dz$  decay time is measured using the transverse displacement
of the $ \Dz $ vertex from the IP and the $ \Dz $
transverse momentum  due to the relative narrowness
of the position distribution of the IP in the transverse ($r$-$\phi$)
plane \cite{Aubert:2001tu}.
Neural networks are used to estimate the boost of signal candidates in
the $r$-$\phi$ plane,
as discussed below.

The pions from $\Dstarp$ decays are relatively soft tracks with $p^{*}_{\pi}<450$~MeV/$c$,
where the asterisk denotes a parameter measured in the $ e^+ e^- $ CM frame.
Charged tracks identified as either a charged $K$ or $e$ candidate are not considered as $\pi$ candidates.
To reject poorly reconstructed tracks, $\pi$ candidates are required to have six or more SVT hits,
with at least 2 hits in each of the $r$-$\phi$ and $z$ views, and at least one hit on the inner three
layers in each of the $r$-$\phi$ and $z$ views.
Pion candidates are refit constraining the tracks to originate from the IP, and are accepted only if
the refit probability is greater than 0.01.
Pion candidates meeting the above criteria are combined with $Ke$ vertex candidates to
form $\Dstarp$ candidates.

A reasonable estimate of a signal candidate's proper decay time cannot be obtained using the partial reconstruction described above and, therefore, the
three orthogonal components of the $\Dz$ CM momentum vector are estimated with three separate JetNet 3.4 \cite{Peterson:1994nk} neural networks. Each
NN has two hidden layers, and is trained and validated with a large sample $(\mathcal{O}(10^{5}))$ of simulated signal events
generated separately from the other MC samples used in the analysis.
The following vector inputs to the NN's are used:
${\bf p}^{*}$($Ke$)
(the momentum of the $Ke$ pair constrained by the vertex fit),
${\bf p}^{*}$($\pi$), and the event thrust vector ${\bf T^*}$ (calculated using all charged and neutral candidates except the $K$ and $e$ candidates).
For simulated signal events,
the distribution of the difference between the true
${\bf p}^{*}$($\Dz$) direction
and the NN output direction is unbiased and Gaussian
with $\sigma \approx 130$~mrad.
The distribution of momentum magnitude differences shows an
uncertainty of $\sigma_{p}/p \approx 10\%$ for a typical signal event.

The transverse momentum of a $\Dz$ candidate and the projections
of the IP and $Ke$ vertex
loci on the $r$-$\phi$ plane are used to calculate a candidate's proper decay time.
The error on the decay time, calculated using only the errors on the IP and $Ke$ vertex,
is typically 0.8$\tau_{\Dz}$, where $\tau_{\Dz}$ is the nominal mean $\Dz$ lifetime.
The contribution of the ${\bf p}^{*}$($\Dz$) estimator to the total decay time uncertainty
is approximately $10\%$ and is ignored.
Poorly reconstructed events, with calculated decay time errors greater than $ 2 \tau_{\Dz}$,
are discarded,
and only events with decay times between
$-12 \tau_{\Dz} $ and $ 15 \tau_{\Dz} $
are retained.
These criteria remove about 7\% of the signal decays.

\begin{figure*}[thb]
\begin{center}
\begin{minipage}{1.0\textwidth}
\epsfig{file=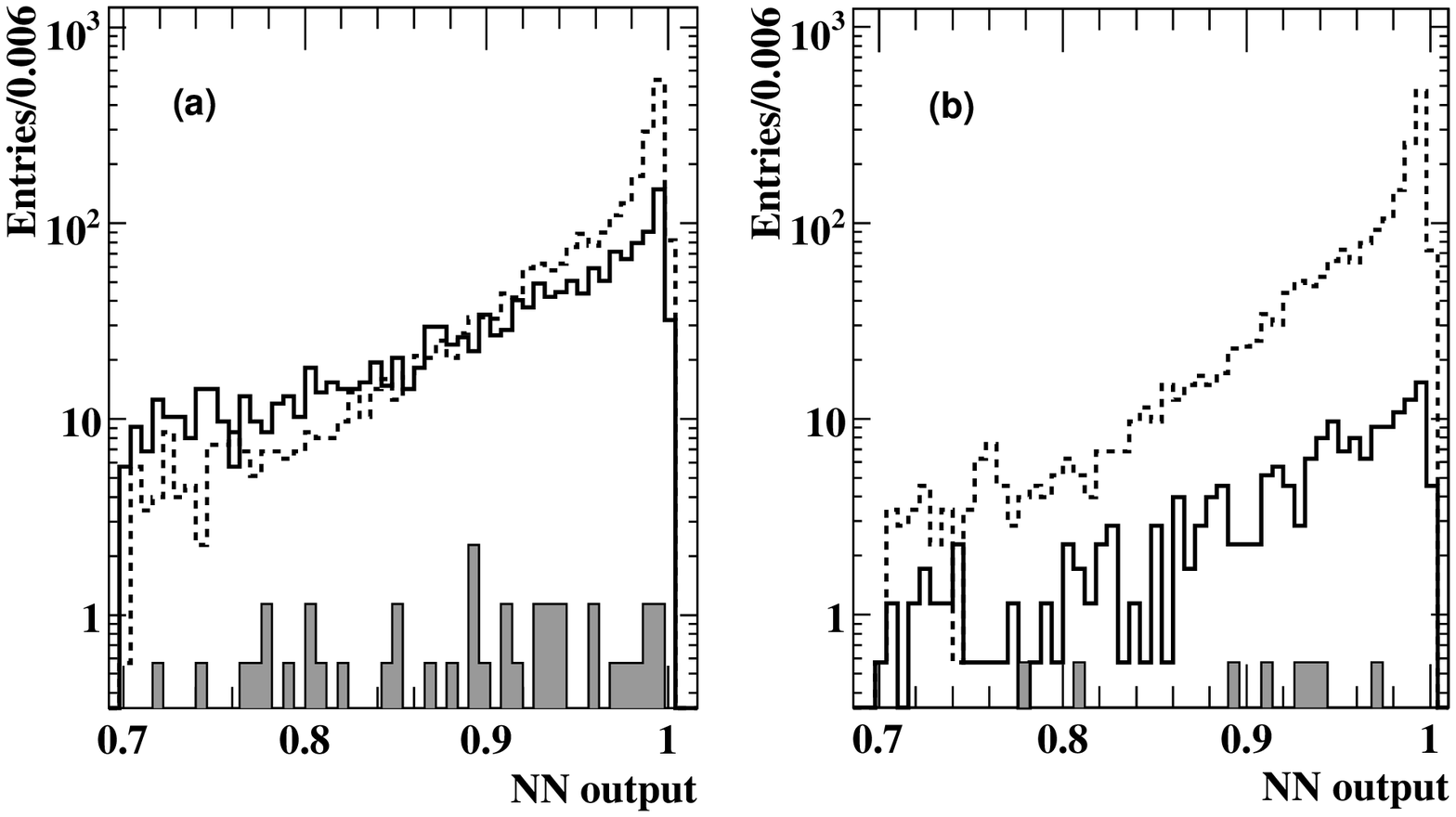,width=0.8\textwidth}
\end{minipage}
\caption[]{\label{fig:nn}
Event selector NN output,
from double-tag luminosity-scaled unbiased  MC, for  events
(a) passing the initial selection of semileptonic side events and
(b) passing the additional semileptonic side selection criteria.
The distributions shown are those for signal candidates (dashed line),
RS backgrounds (solid line), and WS backgrounds (solid fill).
The final event selection requires NN output $>0.9$.
}
\end{center}
\end{figure*}

In addition to the above criteria, events are selected using a neural network trained to
distinguish prompt charm signal from slow pion single-tag WS background events.
The event selector NN uses a five-element input vector:
$p^{*}_{Ke}$, $p^{*}_{\pi}$, $|{\bf T^*}|$, $\theta^{*}
({\bf p}^*_{Ke},{\bf T^*})$,
and $\theta^{*}({\bf p}^*_{K},{\bf p}^*_{e})$ where
$ \theta(\bf a, \bf b ) $ denotes the opening angle
between the vectors $ {\bf a} $ and $ {\bf b} $.
It has a single hidden layer of nine nodes and is also constructed using JetNet~3.4.
Figure~\ref{fig:nn}(a) shows the distribution of NN output for signal candidates, RS
backgrounds and WS backgrounds in double-tag unbiased MC events passing the semileptonic
side event selection criteria given above.
The NN output is required to be greater than 0.9
in order to yield the best statistical sensitivity to WS signal events, assuming a null mixing rate, for the double-tag dataset used here.

%% file: myPubboard/acknowledgements.tex
We are grateful for the 
extraordinary contributions of our \pep2\ colleagues in
achieving the excellent luminosity and machine conditions
that have made this work possible.
The success of this project also relies critically on the 
expertise and dedication of the computing organizations that 
support \babar.
The collaborating institutions wish to thank 
SLAC for its support and the kind hospitality extended to them. 
This work is supported by the
US Department of Energy
and National Science Foundation, the
Natural Sciences and Engineering Research Council (Canada),
the Commissariat \`a l'Energie Atomique and
Institut National de Physique Nucl\'eaire et de Physique des Particules
(France), the
Bundesministerium f\"ur Bildung und Forschung and
Deutsche Forschungsgemeinschaft
(Germany), the
Istituto Nazionale di Fisica Nucleare (Italy),
the Foundation for Fundamental Research on Matter (The Netherlands),
the Research Council of Norway, the
Ministry of Science and Technology of the Russian Federation, 
Ministerio de Educaci\'on y Ciencia (Spain), and the
Particle Physics and Astronomy Research Council (United Kingdom). 
Individuals have received support from 
the Marie-Curie IEF program (European Union) and
the A. P. Sloan Foundation.

%% file: appendix.tex
\appendix
\section{Appendix: Statistical Method for Establishing Confidence Intervals}

To estimate confidence intervals for the number of WS signal events, 
we adapt a method suggested in Ref. \cite{Porter:2003ui}. 
For a true signal rate, $ s $, and background rate, $ b $, in the WS signal region, 
we determine probability density functions (PDF's) for the number of background 
events we should observe in our Monte Carlo simulation, $ n_b $, and the number 
of candidates we should observe in the WS signal region, $ n $. 
We use these to define a global likelihood function for $ s $ and $ b $ which 
depends upon $ n $ and $ n_b $: $  \mathcal{L}(s,b;n,n_b) $. 
Given an observation  $ ( n, n_b $), the central value for $ s $ is that which 
maximizes $  \mathcal{L}(s,b;n,n_b) $. 
The boundaries of confidence intervals for $ s $ 
are then defined by the extremum signal rates in the $ (s,b )$ plane where the logarithm 
of the likelihood function changes by specified values based on those that would 
provide proper frequentist coverage in the limit of high statistics and Gaussian distributions. 
Assuming the PDF's we use are correct, we validate this algorithm by checking the frequentist 
coverage it produces for a range of values of $ s $ and $ b $.

The PDF for $ n_b $ is taken to be
\begin{eqnarray}
& &  P(n_b; b_{\rm MC}) =  \\ \nonumber
  & & N(b_{\rm MC}) \int_0^\infty \frac{x^{n_b}}{n_b!}e^{-x}
    \frac{1}{\sigma(b_{\rm MC})}
    e^{-\frac{1}{2}\frac{(b_{\rm MC}-x)^2}{\sigma(b_{\rm MC})^2}}dx \, .
\label{eqn:bkg}
\end{eqnarray}

In this equation, $ n_b $ and $ b $ are as defined above; $b_{\rm MC} = \alpha b$ is 
the mean number of events expected in a Monte Carlo simulation with $ \alpha $ times 
the luminosity of the data sample, and $\sigma(b_{\rm MC}) = 0.5 \,  b_{\rm MC}$ accounts 
for the 50\% systematic uncertainty in the ratio of background rate in the data and 
background rate in the MC simulation. $N(b_{\rm MC})$ is the normalization such that $\sum_{n_b=0}^{\infty}P(n_b; b_{\rm MC})=1$.

The PDF for the combined observation $ (n,n_b) $ is then taken to be the product 
of $  P(n_b; b_{\rm MC})$ and a purely Poisson term for $ n $:

\begin{eqnarray}
 & &  P(n;n_b) = \\ \nonumber
 & &  P(n_b; b_{\rm MC} =  \alpha b ) \times \frac{(s+b)^n}{n!}e^{-(s+b)} \, .
\label{eqn:total}
\end{eqnarray}
To obtain 90\% (68\%) confidence intervals, we use the following procedure.
\begin{itemize}
\item
We write the global likelihood function for
$ s $ and $ b $ as
\begin{eqnarray}
 & &   \mathcal{L}(s,b;n,n_b)   =    \\ \nonumber
 & &   P(n_b; b_{\rm MC} =   \alpha b ) \times \frac{(s+b)^n}{n!}e^{-(s+b)}\, .
\label{eqn:like}
\end{eqnarray}
\item We find the values  $(\hat s, \hat b)$ for parameters $(s,b)$
 which maximize the likelihood.
 These are $ \hat b = n_b/a $ and $ \hat s = n - \hat b $.
\item With $\mathcal{L}_{max}$  the value of the likelihood at its
  maximum, we obtain a 90\% (68\%) confidence intervals for $s$,
  by finding the
  points in the $(s,b)$ plane where
  \begin{eqnarray}
    \Delta \rm{ln} {\cal L}  & \equiv &
     \rm{ln} {\cal L}_{ \rm max} - \rm{ln} {\cal L}
     {\it (s,b;n,n_b)} \\ \nonumber
  & = & 1.35 \ (0.50).
\label{eqn:deltal}
  \end{eqnarray}
\item We let $s_l$ be the minimum value of $s$ in this set and
  $s_u$ be the maximum value.
  The 90\% (68\%) confidence intervals
  for $s$ are the ranges $(s_l, s_u)$.

\end{itemize}

We have determined the frequentist coverage of this algorithm for many values of 
$ (s,b) $ by using the PDF of Eqn. (\ref{eqn:total}) to generate large samples of $ ( n,n_b) $. 
For any one $ (s, b )$, we consider the ensemble of all $ (n,n_b ) $ generated. 
For each  of these $ (n,n_b) $  we determine whether $s $ is contained in 
the 90\% (68\%) confidence interval defined using the algorithm described above. 
The fraction of all $ (n, n_b) $ containing the true value $s $ is called the coverage. 
The coverage is therefore a function of both $s $ and $ b $ as well as the the 
level (68\% or 90\%). 
As a function of $ s $ between 0 and 10, and for fixed values of $ b $ between 
2.5 and 7 (where 2.85 is the central value ``expected" based on our observation of $ n_b $), 
the coverages we calculate are close to the nominal values. 
In this range, the 68\% intervals provide coverages between 64\% and 72\% with 
the most severe undercoverage observed for $ s < 2 $; 
the 90\% intervals provide coverage between 87\% and 92\%s with the 
most severe undercoverage again observed for  $ s < 2 $. 
The deviations from nominal coverage are relatively small. 
We judge the statistical properties of the quoted intervals to be sufficiently 
accurate for this analysis.